%% file: main.tex
\documentclass{Interspeech2024}
\usepackage{kotex}
\usepackage{pifont}% http://ctan.org/pkg/pifont
\usepackage{multirow}% http://ctan.org/pkg/pifont
\newcommand{\cmark}{\ding{51}}%
\newcommand{\xmark}{\ding{55}}%
\interspeechcameraready

\title{
VoiceTailor: Lightweight Plug-In Adapter for Diffusion-Based Personalized Text-to-Speech
}

\name[affiliation={1}]{Heeseung}{Kim}
\name[affiliation={2}]{Sang-gil}{Lee}
\name[affiliation={1}]{Jiheum}{Yeom}
\name[affiliation={1}]{Che Hyun}{Lee}
\name[affiliation={2}]{Sungwon}{Kim}
\name[affiliation={1,3}]{Sungroh}{Yoon*}

\address{
  $^1$Data Science and AI Lab, ECE, Seoul National University (SNU), Korea\\
  $^2$NVIDIA, USA \\
  $^3$AIIS, ASRI, INMC, ISRC, and Interdisciplinary Program in AI, SNU, Korea}
  
\email{\{gmltmd789, quilava1234, saga1214, sryoon\}@snu.ac.kr, \{sanggill, sungwonk\}@nvidia.com}

\begin{document}

\maketitle
{\let\thefootnote\relax\footnotetext{$*$ Corresponding Author}}
\begin{abstract}
    We propose VoiceTailor, a parameter-efficient speaker-adaptive text-to-speech (TTS) system, by equipping a pre-trained diffusion-based TTS model with a personalized adapter. VoiceTailor identifies pivotal modules that benefit from the adapter based on a weight change ratio analysis.
    We utilize Low-Rank Adaptation (LoRA) as a parameter-efficient adaptation method and incorporate the adapter into pivotal modules of the pre-trained diffusion decoder.
    To achieve powerful adaptation performance with few parameters, we explore various guidance techniques for speaker adaptation and investigate the best strategies to strengthen speaker information. 
    VoiceTailor demonstrates comparable speaker adaptation performance to existing adaptive TTS models by fine-tuning only 0.25\% of the total parameters. 
    VoiceTailor shows strong robustness when adapting to a wide range of real-world speakers, as shown in the demo\footnote{{Demo: \href{https://voicetailor.github.io/}{https://voicetailor.github.io/}}}.
\end{abstract}

\input{sections/introduction}
\input{sections/method}
\input{sections/experiments}
\input{sections/conclusion}

\section{Acknowledgements}
This work was supported by Samsung Electronics (IO221213-04119-01), Institute of Information \& Communications Technology Planning \& Evaluation (IITP) grant funded by the Korea government (MSIT) (RS-2021-II211343: AI Graduate School
Program, SNU, RS-2022-II220959), National Research Foundation of Korea grant funded by MSIT (2022R1A3B1077720), and the BK21 FOUR program of the Education and Research Program for Future ICT Pioneers, SNU in 2024.

\bibliographystyle{IEEEtran}
\bibliography{main}

\end{document}

%% file: sections/introduction.tex
\section{Introduction}

Recent advancements in deep generative models have led to improvements in adaptive text-to-speech (TTS), enabling models to generate a target speaker's voice from a given transcript and reference speech \cite{pmlr-v162-casanova22a, kim2022guidedtts, le2023voicebox}.
Zero-shot approach \cite{pmlr-v162-casanova22a, le2023voicebox, wang2023neural, kim2023pflow, shen2024naturalspeech} for adaptive TTS eliminates the need for extra fine-tuning on reference audio for speaker adaptation.
Despite its advantage of no further training, this approach generally requires large speech corpus during training to achieve high speaker similarity, and is comparatively less robust against unique out-of-distribution voices commonly encountered in real-world scenarios. 

One-shot approach, an alternative type of adaptive TTS, constructs personalized TTS by fine-tuning pre-trained multi-speaker TTS models with few reference speeches of target speaker \cite{pmlr-v162-casanova22a, DBLP:journals/corr/abs-2005-05642, Moss2020BOFFINTF, hsieh23_interspeech, NEURIPS2018_4559912e, chen2021adaspeech, 9414872}. 
To efficiently adapt to the target speaker, several studies fine-tuned a subset of the model's parameters \cite{DBLP:journals/corr/abs-2005-05642, Moss2020BOFFINTF, NEURIPS2018_4559912e, chen2021adaspeech, 9414872}, or leveraged adapter-based fine-tuning techniques \cite{hsieh23_interspeech} such as Low-Rank Adaptation (LoRA) \cite{hu2022lora} or prefix-tuning \cite{li-liang-2021-prefix}, which only fine-tune the parameters of newly integrated adapters.
However, these works often fail to generate speech with high speaker similarity due to the limitations of the generative models used as decoder and typically require more than a minute of speech data for fine-tuning.

Recently, inspired by successes of diffusion-based generative model \cite{DDPM} on fine-tuning-based personalized generation tasks \cite{Ruiz_2023_CVPR}, diffusion-based one-shot TTS models have been proposed \cite{kim2022guidedtts, kim23k_interspeech}. 
They leverage the diffusion model's adaptation performance to achieve high speaker similarity in personalized TTS task with as short as 5 to 10 seconds of reference speech. 
However, in contrast to other one-shot approaches, these works fine-tune all model parameters, resulting in parameter inefficiency.

In this work, we introduce VoiceTailor, a parameter-efficient adaptive TTS model that requires fine-tuning only a subset of parameters from a diffusion-based pre-trained TTS model.
We utilize a diffusion-based pre-trained TTS model and adopt a fine-tuning methodology following UnitSpeech \cite{kim23k_interspeech}. 
Inspired by the approaches in \cite{kumari2022customdiffusion, 10.5555/3495724.3497056}, we analyze the change ratio in the weights of each module in the model before and after fine-tuning and identify that attention modules play a crucial role in speaker adaptation. 
Based on this observation, VoiceTailor carefully integrates LoRA into the effective attention modules in the model and fine-tunes only the injected low-rank matrices for adaptation.

We demonstrate that VoiceTailor achieves speaker adaptation performance comparable to the fully fine-tuned one-shot baseline by plugging in the small adapter with $0.25\%$ of the total parameters of the pre-trained model, which occupies approximately 1.3 MB of storage space.
In addition, we systematically analyze the impact of various design choices and hyperparameters during the parameter-efficient adaptation stage.
Furthermore, we investigate the best strategy from various guidance techniques in the inference stage. 
We illustrate VoiceTailor's robust performance in real-world scenarios by presenting a variety of samples, including those adapted for real-world speakers, on our demo page. Our contributions are as follows:

\begin{itemize}
\item To the best of our knowledge, this is the first work that systematically incorporates LoRA for diffusion-based speaker adaptive TTS that achieves high speaker similarity.
\item VoiceTailor significantly reduces cost of adapting TTS to new speaker using 10 seconds of untranscribed speech with approximately 15 seconds of training time on a single GPU by utilizing $0.25\%$ of the model parameters.
\item We compare and analyze various methods to enhance speaker information using LoRA modules and speaker classifier-free guidance and investigate the optimal strategy.
\end{itemize}

%% file: sections/method.tex
\section{Method}
\begin{figure*}[h]
    \centering
    \includegraphics[width=0.88\linewidth]{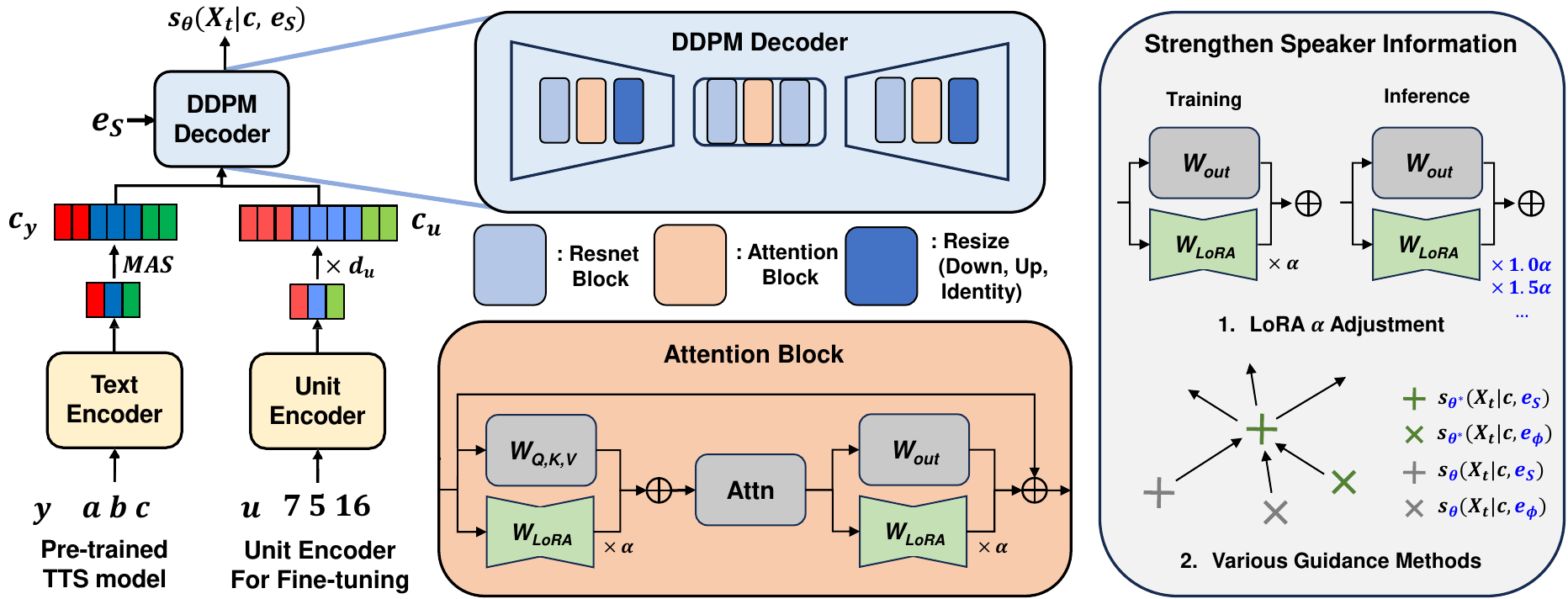}
    \caption{An overview of VoiceTailor depicting the LoRA adapters and techniques for strengthening speaker information.}
    \label{fig1}
    \vskip -0.2in
\end{figure*}

We introduce VoiceTailor, a personalized TTS model utilizing LoRA to address the parameter inefficiency prevalent in existing diffusion-based one-shot TTS approaches. 
VoiceTailor captures the target speaker's characteristics through LoRA fine-tuning and a speaker embedding extracted from a reference audio. We conduct a weight change ratio analysis of an existing model, UnitSpeech \cite{kim23k_interspeech}, and explore various methodologies to enhance the speaker information. Through careful injection of LoRA weights from our analysis and selecting the optimal strategy for guidance technique, VoiceTailor achieves personalized TTS by fine-tuning as few as $0.25\%$ of the model's total parameters. 
A detailed overview of VoiceTailor is illustrated in Figure \ref{fig1}. 
UnitSpeech, our baseline model for the one-shot approach, is specified in Section \ref{method:unitspeech}. We describe details of the fine-tuning process using LoRA in Section \ref{method:peft}. 
We introduce several strategies to strengthen the target speaker information when synthesizing personalized speech in Section \ref{method:guidance}.

\subsection{UnitSpeech}
\label{method:unitspeech}
In this work, we employ UnitSpeech \cite{kim23k_interspeech}, an adaptive speech synthesis model with powerful personalization capabilities, serving as the foundation for our one-shot TTS approach.
UnitSpeech introduces a method to construct a personalized TTS model by fine-tuning a pre-trained, multi-speaker, diffusion-based TTS model with a short untranscribed speech sample.

The multi-speaker diffusion-based TTS model in UnitSpeech is based on Grad-TTS \cite{Grad-TTS}, which first defines a forward process that converts a mel-spectrogram $X_0$ to Gaussian noise $X_T\sim N(0,I)$. 
The forward process is defined using the pre-defined noise schedule $\beta_t$ and the Wiener process $W_t$. The noisy mel-spectrogram $X_t$ at timestep $t\in[0,T]$ in the forward process is computed as follows:
\begin{align}   
    \label{forward process}
        dX_t &= -\frac{1}{2}X_t\beta_tdt+\sqrt{\beta_t}dW_t,\quad t\in[0,1],\\
        X_t&= \sqrt{{\rm e}^{-\int_0^t\beta_{s}ds}} X_0 + \sqrt{1 - {\rm e}^{-\int_0^t\beta_{s}ds}} \epsilon_t.
\end{align}
Here, $\epsilon_t$ is the noise sampled from the standard normal distribution. 

To sample the mel-spectrogram along the reverse trajectory of the previously defined process, it is necessary to utilize a score $s(X_t|c_y,e_S)$ that is conditioned on the text encoder output $c_y$ and the speaker embedding $e_S$ extracted from the pre-trained speaker encoder.
UnitSpeech's diffusion-based decoder $\theta$ is trained to predict the conditional score $s_{\theta}(X_t|c_y,e_S)$. 
The loss function for decoder pre-training and the formula of using the predicted score for sampling are as follows:
\begin{align}   
    \label{reverse process}        
        L&={\mathbb{E}_{t,X_0,\epsilon_t}[\lVert(\sqrt{1 - {\rm e}^{-\int_0^t\beta_{s}ds}}s_\theta(X_t|c_y,e_S)+\epsilon_t\rVert_2^2]}], \\       
        X&_{t-\Delta{t}} = X_t + \beta_t(\frac{1}{2}X_t + s_\theta(X_t|c_y, e_S))\Delta{t} + \sqrt{\beta_t\Delta{t}}z_t,
\end{align}
where $z_t\sim N(0,I)$ is Gaussian noise.

UnitSpeech introduces a unit encoder to fine-tune the pre-trained diffusion decoder with untranscribed speech, eliminating the need for text input during the speaker adaptation process. 
The unit encoder is designed to replace the text encoder by receiving acoustic units (\textit{i.e.,} self-supervised speech representations containing phonetic information \cite{polyak21_interspeech}) as input.
By substituting the text encoder with this pluggable unit encoder and training it with the same objective as the pre-trained decoder, UnitSpeech can receive unit inputs in addition to text inputs.
This approach enables speaker adaptation by fine-tuning the decoder with the reference audio and its corresponding unit.

UnitSpeech integrates classifier-free guidance \cite{ho2021classifierfree}, a method for enhancing conditioning information in diffusion models, into the text encoder output $c_y$ for accurate pronunciation. 
Unlike UnitSpeech, which solely applies classifier-free guidance to text conditions, we extend this approach to speaker embeddings $e_S$ as well.
While pre-training the multi-speaker TTS model, we introduce a learnable unconditional embedding $e_\phi$ and substitute $e_S$ with $e_\phi$ with a probability of $25\%$. 
The resulting unconditional score obtained with $e_\phi$ is then utilized for speaker classifier-free guidance, as detailed in Section \ref{method:guidance}.
  
\subsection{Parameter-Efficient Speaker Adaptation}
\label{method:peft}
To address the inefficiency of fine-tuning all parameters during speaker adaptation, we incorporate LoRA \cite{hu2022lora}, a parameter-efficient adaptation technique.
LoRA is a method that allows fine-tuning of the linear layer's weight matrix by combining trainable low-rank decomposed matrices.
Given a pre-trained weight $W\in\mathbb{R}^{d\times k}$ of the linear layer, LoRA augments it with $W + \alpha \cdot \Delta W= W + \alpha \cdot BA$, where the parameters $\Delta W:=W_{LoRA}$ are fine-tuned with $W$ being frozen.
Here,  $B\in \mathbb{R}^{d\times r}, A\in \mathbb{R}^{r\times k}$, $\alpha$ is the scaling factor of the adapter matrices, and $r$ represents the rank.
By using a significantly smaller value for the rank $r$ compared to the dimensions $d,k$ of the original matrix ($r \ll d,k$), LoRA enables adaptation with orders of magnitude fewer parameters. We denote the pre-trained model's parameters as $\theta$, and the parameters of the model with the fine-tuned adapter ($W_{LoRA}$) as $\theta^*$.

Inspired by \cite{kumari2022customdiffusion, 10.5555/3495724.3497056}, we first conduct speaker adaptation by fine-tuning all decoder parameters using UnitSpeech to explore which modules play a pivotal role in speaker adaptation.
We measure the weight change ratio $||\theta^*_i - \theta_i|| / ||\theta_i||$ for each module $\theta_i$ before and after fine-tuning. 
Considering the prevalent application of LoRA to attention modules \cite{hu2022lora, kumari2022customdiffusion}, we measure the average change ratios of weight in the attention module and other modules within UnitSpeech's diffusion decoder, obtaining values of $0.0282$ and $0.0050$, respectively.
These results confirm that, similar to \cite{kumari2022customdiffusion}, the attention module is crucial in adaptation for one-shot diffusion-based TTS models.
Consequently, we inject LoRA into the attention module and optimize only these parameters for speaker adaptation.
During this fine-tuning process, we use the same objective used for DDPM decoder pre-training in UnitSpeech, as specified in Eq. \ref{reverse process}. 

\subsection{Speaker Information Strengthening Strategies}
\label{method:guidance}

The fine-tuned adapter, combined with the pre-trained multi-speaker TTS model, enables us to construct personalized TTS for the target speaker. 
In VoiceTailor, the speaker information is provided in two forms: the speaker embedding ($e_S$) and the pluggable LoRA weights ($W_{LoRA}$).
To mitigate degradation in speaker adaptation performance due to decreased parameters, we explore various approaches for sampling to strengthen the target speaker's information. 
We consider adjusting the scaling factor $\alpha$ of LoRA to a value greater than what is used during fine-tuning, and applying classifier-free guidance to both forms of information.

\noindent\textbf{Adjustment of LoRA scaling factor} 
$\alpha$ controls the intensity with which the adapter is added to the pre-trained model for speaker adaptation.
By using a larger $\alpha$ during generation than the one used during training, we aim to provide stronger speaker information contained within the low-rank adapter.

\noindent\textbf{Classifier-free guidance} 
As there are two sources of speaker information ($e_S$ and $W_{LoRA}$), we consider classifier-free guidance for each source. Given the score of fine-tuned model $s_{\theta^*}(X_t|c,e_S)$, we consider $3$ candidates for the unconditional score $s_{uncon}$:
\begin{enumerate}
\item $s_{\theta^*}(X_t|c,e_\phi)$ can be obtained by replacing $e_S$ with the unconditional embedding $e_\phi$ while maintaining the speaker information provided by $W_{LoRA}$.
\item $s_\theta(X_t|c,e_S)$ can be obtained from the pre-trained model $\theta$ by removing $W_{LoRA}$ and keeping $e_S$ as input. 
\item $s_\theta(X_t|c,e_\phi)$ can also be used as $s_{uncon}$ which lacks all speaker information from $e_S$ and $W_{LoRA}$.
\end{enumerate}

The modified score $\hat{s}$ is calculated by applying classifier-free guidance with the above unconditional scores as follows:
\begin{equation} 
    \label{classifier-free guidance}
        \hat{s}_{\theta^*}(X_t|c,e_S) = s_{\theta^*}(X_t|c,e_S) + \gamma_S \cdot (s_{\theta^*}(X_t|c,e_S) - s_{uncon}).
\end{equation}
Here, $\gamma_S$ represents the gradient scale which determines the intensity of the additional speaker information.

We perform TTS with the $4$ methods (adjusting $\alpha$ and $3$ candidates for $s_{uncon}$) described above and observe that methods other than applying classifier-free guidance with $s_{uncon}=s_{\theta^*}(X_t|c,e_\phi)$ lead to detrimental performance in speaker adaptation.
Therefore, when generating samples with VoiceTailor, we adopt using $s_{uncon}=s_{\theta^*}(X_t|c,e_\phi)$ as our final method. 
The related results and analysis are presented in Section \ref{analysis}.

%% file: sections/experiments.tex
\section{Experiments}
\subsection{Experimental Setup}
\subsubsection{Datasets}
Similar to UnitSpeech, we train a multi-speaker diffusion-based TTS model using the LibriTTS dataset \cite{zen19_interspeech} which comprises $585$ hours of speech-text data across $2,456$ speakers.
We employ the same speaker encoder as UnitSpeech trained on VoxCeleb 2 \cite{Voxceleb2}. 
For evaluation purpose, we select 10 speakers from the LibriTTS \texttt{test-clean} subset choosing one reference audio for each speaker which is identical to the reference audio used in YourTTS \cite{pmlr-v162-casanova22a}. 
We select 5 random samples for each speaker, resulting in a total of 50 samples for evaluation.

\begin{table*}
\caption{Results of one/zero-shot adaptive TTS models including mean opinion score (MOS), character error rate (CER), and speaker similarity mean opinion score (SMOS) with $95\%$ CI. The Amount of Dataset denotes the volume of data used to train the multi-speaker TTS model, measured in hours. \# Params refers to the number of parameters utilized for fine-tuning / the total number of parameters.}
\vskip -0.2in
\label{table:main}
\footnotesize
\begin{center}
\begin{tabular}{l|c|c|c|ccc}
\toprule
\textbf{Method} & \textbf{Amount of Dataset} & \textbf{Fine-tuning}  & \textbf{\# Params} & \textbf{5-scale MOS}       & \textbf{CER (\%)} & \textbf{5-scale SMOS}      \\ \midrule
Ground Truth    & -                & -                     & -              & $4.36\pm0.05$ & $0.70$              & $4.38\pm0.06$ \\
Mel + BigVGAN \cite{lee2023bigvgan}   & -                & -                     & -               & $4.23\pm0.07$ & $0.73$               & $4.16\pm0.08$ \\ \midrule
VoiceTailor    & $\approx 585$ hrs  & \cmark &  $0.311M$ / $127M$          & $4.19\pm0.07$ & $1.33$               & $4.06\pm0.09$ \\
UnitSpeech \cite{kim23k_interspeech}     & $\approx 585$ hrs  & \cmark &       $119M$ / $127M$                             & $4.13\pm0.07$ & $1.24$               & $4.08\pm0.10$ \\ \midrule
XTTS \textit{v2}         &       $> 16,000$ hrs           & \xmark &       $0$ / $467M$                            & $4.14\pm0.08$ & $1.18$               & $3.85\pm0.11$ \\
YourTTS \cite{pmlr-v162-casanova22a}        &     $\approx 474$ hrs            & \xmark &                 $0$ / $87M$                   & $3.87\pm0.09$ & $2.78$               & $3.67\pm0.10$ \\
\bottomrule
\end{tabular}
\vskip -0.5in
\end{center}
\vskip -0.3in
\end{table*}

\subsubsection{Training and Fine-tuning Details}
For training the multi-speaker TTS model, we adhere to the UnitSpeech architecture but introduce a learnable unconditional speaker embedding $e_\phi$ during training to facilitate speaker classifier-free guidance. 
Training procedures are consistent with those of UnitSpeech.
For speaker adaptation, we fine-tune $W_{LoRA}$ for $500$ iterations using the Adam optimizer \cite{DBLP:journals/corr/KingmaB14} at a learning rate of $10^{-4}$, which takes approximately 15 seconds using a single NVIDIA A100 GPU. 
Compared to UnitSpeech, VoiceTailor performs fine-tuning with a higher learning rate due to its significantly fewer parameters for adaptation.
We set the LoRA rank $r$ and scaling factor $\alpha$ to 16 and 8, respectively. 
By setting $r=16$, we fine-tune only $311K$ of the total $127M$ parameters of the model, which corresponds to 0.25\% of the total and amounts to a size of 1.3 MB in storage.
 
\subsubsection{Evaluation}
During evaluation, we select UnitSpeech as our one-shot baseline.
Additionally, we choose YourTTS \cite{pmlr-v162-casanova22a} as the zero-shot TTS baseline which is trained on a similar scale of speech data, and XTTS $v2$, a powerful open-source zero-shot TTS model known to be trained on over 16,000 hours of data.
For the vocoder, we use the official checkpoint of BigVGAN \cite{lee2023bigvgan}.
During sampling, we use the same LoRA scale $\alpha=8$ as used in training, set the speaker gradient scale $\gamma_S=1$, and use step size $\Delta t=0.02$. All samples are resampled to 16kHz and are normalized to $-27$dB for a fair comparison.

We utilize a test set of 50 sentences to evaluate the performance of VoiceTailor.
We evaluate subjective audio quality and naturalness of generated samples through a 5-scale mean opinion score (MOS) and the speaker similarity with a 5-scale speaker similarity mean opinion score (SMOS). We also measure objective metrics with the speaker encoder cosine similarity (SECS), and the character error rate (CER) for evaluating pronunciation accuracy.
The MOS and SMOS assessments are conducted using MTurk, while the SECS and CER measurements employ Resemblyzer package's speaker encoder \cite{resemblyzer} and CTC-based Conformer \cite{gulati20_interspeech}, respectively. 
Following UnitSpeech, we generate each sentence $5$ times for the SECS and CER measurements and average the values.

\begin{table}
\caption{CER and SECS results for design choices. The final setup marked in \textbf{bold}. \textit{``attn + others''}: injection of adapters to all linear layers in addition to the attention modules.}
\vskip -0.2in
\label{table:ablation_finetuned}
\footnotesize
\begin{center}
\begin{tabular}{cc|cc}
\toprule
\multicolumn{2}{c|}{\textbf{}}                                                 & \textbf{CER (\%)} & \textbf{SECS} \\ \midrule
\multicolumn{1}{c|}{\multirow{2}{*}{LoRA Modules}}           & \textbf{\textit{attn}}          & \bm{$1.33$}           & \bm{$0.942$}         \\
\multicolumn{1}{c|}{}                                          & \textit{attn + others} & $1.39$            & $0.941$         \\ \midrule
\multicolumn{1}{c|}{\multirow{5}{*}{\begin{tabular}[c]{@{}c@{}}LoRA Rank $r$ \\ (\# Trainable Params)\end{tabular}}}                 & $2\ (39K)$             & $1.37$            & $0.939$         \\
\multicolumn{1}{c|}{}                                          & $4\ (78K)$             & $1.41$            & $0.939$         \\
\multicolumn{1}{c|}{}                                          & $8\ (156K)$             & $1.47$            & $0.940$        \\
\multicolumn{1}{c|}{}                                          & \bm{$16\ (311K)$}             & \bm{$1.33$}            & \bm{$0.942$}        \\
\multicolumn{1}{c|}{}                                          & $32\ (623K)$             & $1.35$            & $0.941$       \\  \midrule
\multicolumn{1}{c|}{\multirow{4}{*}{LoRA Scale $\alpha$}}      & $1$             & $1.30$            & $0.926$         \\
\multicolumn{1}{c|}{}                                          & $2$             & $1.24$            & $0.937$         \\
\multicolumn{1}{c|}{}                                          & $4$             & $1.29$            & $0.939$         \\
\multicolumn{1}{c|}{}                                          & \bm{$8$}             & \bm{$1.33$}            & \bm{$0.942$}         \\ \midrule
\multicolumn{1}{c|}{\multirow{4}{*}{LR / \# Iters}} & $2\cdot 10^{-5}$ / $500$    & $1.23$            & $0.912$         \\
\multicolumn{1}{c|}{}                                          & $2\cdot 10^{-5}$ / $2000$   & $1.25$            & $0.942$         \\
\multicolumn{1}{c|}{}                                          & \bm{$10^{-4}$} / \bm{$500$}    & \bm{$1.33$}            & \bm{$0.942$}         \\
\multicolumn{1}{c|}{}                                          & $10^{-4}$ / $2000$   & $1.49$            & $0.942$      
   \\  
\bottomrule
\end{tabular}
\vskip -0.5in
\end{center}
\vskip -0.35in
\end{table}

\subsection{Results}
\subsubsection{Model Comparison}
We conduct comparative evaluations of our model against various baselines in adaptive text-to-speech, with the results detailed in Table \ref{table:main}. 
As observed in Table \ref{table:main}, VoiceTailor is capable of synthesizing high-quality speech comparable or superior to the baselines, with accurate pronunciation accuracy. 

From the SMOS results measuring speaker similarity, we find that VoiceTailor matches UnitSpeech and exhibits superior adaptation performance to YourTTS, a zero-shot approach utilizing similar amounts of data ($p<0.01$ in the Wilcoxon signed-rank test). Despite using significantly less data, VoiceTailor outperforms XTTS $v2$ in SMOS ($p<0.05$), a zero-shot TTS model trained on vastly larger datasets with larger model size. Notably, fine-tuning only $0.25\%$ of parameters results in comparable speaker similarity to UnitSpeech which fine-tunes the whole parameters, highlighting the efficiency over existing diffusion-based one-shot TTS models in the adaptation.

% We provide samples from each model on our demo page. 
% We also provide various generated samples with real-world reference clips to demonstrate the robust performance of parameter-efficient fine-tuning in real-world scenarios. 
% \textbf{We highly recommend readers to listen to the samples in our demo page.}

\subsubsection{Analysis}
\label{analysis}
We investigate the impact of various factors that could affect LoRA-based speaker adaptation. 
Results on design choices during the fine-tuning process are in Table \ref{table:ablation_finetuned}, while results related to the speaker information strengthening methodology during inference are in Table \ref{table:ablation_inference}.

\noindent\textbf{Parameter-Efficient Fine-Tuning} 
As in Table \ref{table:ablation_finetuned}, additionally injecting trainable low-rank matrices into linear layers other than attention ($attn+others$) does not improve pronunciation accuracy and speaker similarity. 
This aligns with the observation in Section \ref{method:peft} that attention modules are crucial for speaker adaptation. 
Unlike UnitSpeech, which uses a learning rate of $2\cdot10^{-5}$, VoiceTailor requires a higher learning rate due to its adaptation with significantly fewer parameters. 
The choice of $\alpha$ for determining the scale of $W_{LoRA}$ during fine-tuning indicates that comparable speaker similarities can be achieved as long as it is not defined as a small value (\textit{e.g.,} $\alpha=1$). 
Even an extremely small LoRA rank ($r=2$) degrades SECS slightly, suggesting that VoiceTailor can perform speaker adaptation with as few as $39K$ parameters (0.18 MB), should minor performance losses be deemed acceptable for significant parameter efficiency. 

\begin{table}
\caption{CER and SECS results for speaker information strengthening techniques for sampling. The final setup marked in \textbf{bold}. ``$2.0 \cdot \alpha$'': doubles $\alpha$ used for training at inference.}
\vskip -0.2in
\label{table:ablation_inference}
\footnotesize
\begin{center}
\begin{tabular}{c|c|cc}
\toprule
\multicolumn{2}{c|}{\textbf{}}  & \textbf{CER (\%)} & \textbf{SECS} \\ \midrule
w/o strengthening                                                                                          & -                    & $1.25$            & $0.934$       \\ \midrule
LoRA scale (sampling)                                                                                                         & $2.0\cdot\alpha$     & $7.46$            & $0.863$       \\ \midrule
\multirow{2}{*}{\begin{tabular}[c]{@{}c@{}}Gradient scale $\gamma_S$\\ $(s_{uncon} = s_{\theta^*}(X_t|c,e_\phi))$\end{tabular}} & \bm{$1.0$}                & \bm{$1.33$}            & \bm{$0.942$}       \\
                                                                                                                              & $2.0$                & $1.40$            & $0.941$       \\ \midrule
\multirow{2}{*}{\begin{tabular}[c]{@{}c@{}}Gradient scale $\gamma_S$\\ $(s_{uncon} = s_{\theta}(X_t|c,e_S))$\end{tabular}}      & $1.0$                & $1.38$            & $0.918$       \\
                                                                                                                              & $2.0$                & $1.40$            & $0.895$       \\ \midrule
\multirow{2}{*}{\begin{tabular}[c]{@{}c@{}}Gradient scale $\gamma_S$\\ $(s_{uncon} = s_{\theta}(X_t|c,e_\phi))$\end{tabular}}   & $1.0$                & $1.26$            & $0.929$       \\
                                                                                                                              & $2.0$                & $1.46$            & $0.916$        
\\
\bottomrule
\end{tabular}
\vskip -0.5in
\end{center}
\vskip -0.35in
\end{table}

\noindent\textbf{Speaker Information Strengthening Methods} 
We explore various techniques to strengthen the speaker information in the sampling procedure. 
The quantitative results presented in Table \ref{table:ablation_inference} show that except for classifier-free guidance based on the speaker embedding $e_S$ $(s_{uncon} = s_{\theta^*}(X_t|c,e_\phi))$, other techniques deteriorate speaker adaptation performance. 
For example, elevating the LoRA scaling factor $\alpha$ above the value used for fine-tuning degrades both CER and SECS on a large scale. 
Thus, we only apply speaker embedding guidance with $\gamma_S=1$.

%% file: sections/conclusion.tex
\section{Conclusion}
We introduce VoiceTailor which is capable of performing high-quality personalized TTS with a pluggable and small personalized adapter.
VoiceTailor maximizes parameter efficiency by careful injection of LoRA into pivotal modules for speaker adaptation based on the weight change ratio analysis, alongside exploring various guidance techniques to strengthen the speaker information.
Consequently, we demonstrate that VoiceTailor is able to achieve performance comparable to fully fine-tuned adaptive TTS baselines with only $0.25\%$ of the parameters and further show its robustness in real-world scenarios.

We believe that VoiceTailor will reduce the burden of building a personalized TTS system to support numerous new speakers efficiently.
Nonetheless, there is room for further improvements in our parameter-efficient speaker adaptation.
Future directions could include exploring methodologies for conducting speaker adaptation with even fewer parameters without performance degradation and extending the method to other adaptive speech synthesis tasks, such as any-to-any voice conversion.